\documentclass[12pt]{iopart}
\usepackage{iopams, graphics, epsfig}  
\begin{document}

\title[Van der Waals quintessence]{Constraining Van der Waals quintessence by observations}

\author{S. Capozziello$^{1,2}$\footnote[1]{Present address\,: Dipartimento di Scienze Fisiche, Universit\`a di Napoli, Compl. Univ. di Monte S. Angelo, Edificio G, Via Cinthia, 80121 - Napoli, Italy}, V.F. Cardone$^{1,2}$\footnote[2]{Corresponding author\,: {\tt winny@na.infn.it}}, S. Carloni$^{3}$, S. De Martino$^{1,4}$, M. Falanga$^{1,4}$, A. Troisi$^{1,2}$ and M. Bruni$^{5}$}

\address{$^1$Dipartimento di Fisica ``E.R. Caianiello'', Universit\`a di Salerno, via S. Allende, 84081 - Baronissi (Salerno), Italy}

\address{$^2$INFN, Sezione di Napoli, Gruppo Collegato di Salerno, via S. Allende, 84081 - Baronissi (Salerno), Italy}

\address{$^3$Department of Mathematics and Applied Mathematics, University of Cape Town, Rondebosch 7700, Cape Town, South Africa}

\address{$^4$INFM, Sezione di Salerno, Gruppo Coll. di Salerno, via S. Allende, 84081 - Baronissi (Salerno), Italy}

\address{$^5$Institute of Cosmology and Gravitation, Portsmouth University, Mercantile House, PO1 2EG - Portsmouth, United Kingdom}

\begin{abstract}
We discuss an alternative approach to quintessence modifying the usual equation of state of the cosmological fluid in order to see if going further than the approximation of perfect fluid allows to better reproduce the available data. We consider a cosmological model comprising only two fluids, namely baryons (modelled as dust) and dark matter with a Van der Waals equation of state. First, the general features of the model are presented and then the evolution of the energy density, the Hubble parameter and the scale factor are determined showing that it is possible to obtain accelerated expansion choosing suitably the model parameters. We use the the data on the dimensionless coordinate distances to Type Ia supernovae and distant radio galaxies to see whether Van der Waals quintessence is viable to explain dark energy and to constrain its parameters. We then compare the model predictions with the estimated age of the universe and the position of the first three peaks of the anisotropy spectrum of the cosmic microwave background radiation. 
\end{abstract}


\section{Introduction}

In the last few years an increasing bulk of data have been accumulated favouring the scenario of a spatially flat universe dominated by some form of dark energy. A first strong evidence came from the Hubble diagram of type Ia supernovae (hereafter SNeIa) that turned out to be best fitted by spatially flat accelerating cosmological models including a non trivial component \cite{SNeIa}. On the other hand, the results from the observed first and second peak in the cosmic microwave background radiation (hereafter CMBR) spectrum strongly suggested that the geometry of the universe is spatially flat \cite{Boom,CMBR}. When combined with the data on the matter density parameter $\Omega_M$, these results lead to the conclusion that the contribution $\Omega_X$ of the dark energy is the dominant one, being $(\Omega_M, \Omega_X) \simeq (0.3, 0.7)$. This picture of the universe has been further strengthened by an increasing sample of high redshift SNeIa \cite{NewSNeIa,Riess04} and most precise and extended measurements of the CMBR spectrum \cite{WMAP,VSA}. 

After the discovery of these evidences of a spatially flat and accelerating universe, an overwhelming flood of papers, presenting a great variety of models for the explanation of this phenomenon, has appeared. The simplest explanation is the cosmological constant \cite{Lambda} which is able to fit the SNeIa data with good confidence, but it is also plagued by many problems on different scales. This situation has strongly encouraged the search for alternative approaches which now ranges from minimal coupled scalar fields, to strings and anthropic principle (see, e.g., \cite{QuintRev} and references therein).

In a recent work \cite{CdMF}, some of us have introduced a new approach to the problem modifying the usual equation of state of the cosmological fluid in order to see whether going further than the standard  approximation of cosmological perfect fluid allows to  reproduce the available data. The authors have considered a standard cosmology with a Van der Waals equation of state for matter without any other kind of energy source. This approach is extremely "natural" and "obvious" since it starts from the consideration that the universe, in its evolution, is not always well described by a perfect fluid in the forms of radiation or non\,-\,interacting dust. In our opinion, before adding an exotic and mysterious dark energy into the cosmic pie, it is worth wondering whether the observed acceleration of the universe can be implemented using the minimal number of fluids\footnote{Following Newton {\it ``Hypotheses non fingo''}.}. The price to pay is taking into account a more complicated equation of state and motivate it physically. From elementary thermodynamics, we know that a real fluid is never perfect \cite{Rowlinson}. Moreover, it is also well known that the perfect fluid equation of state $p = \gamma \rho$ with $\dot{\gamma} = 0$ is just a rough approximation of cosmic epochs capable of describing stationary situations where phase transitions (e.g., from radiation dominated to dust dominated regions) are not considered \cite{E93+EvE98}. On the other hand, the only thing we know about dark energy is that it gives rise to an accelerated expansion, but there are no hints about its nature. Due to this situation, it is worth asking whether dark energy is indeed needed or, on the contrary, the observed acceleration is driven by standard dark matter provided that its equation of state is more realistically treated. 

A Van der Waals fluid could be a first step toward the goal to get a whole dynamics where {\it 1.)} only observed fluids are taken into account, {\it 2.)} phase transitions occur in the framework of the same evolution, {\it 3.)} accelerated and decelerated periods depend on the relative values of the parameters of the state equation with respect to the pressure and matter energy density which are functions of time. It is worth noting that a similar approach has yet been explored in \cite{Kr03} where the author considers a mixture of two fluids, using the perfect gas equation of state for the matter and the Van der Waals one for the dark energy. However, our approach is radically different since the model we consider is made out of matter only. Is is worth noting that the term {\it matter} usually refers to both baryons and dark matter and both these substances are described by the same equation of state. Actually, we have a direct knowledge of the properties of baryons only and indeed their equation of state is well described (on cosmological scales) by the dust approximation. On the opposite, the nature of the dark matter is still completely unknown so that, a priori, nothing prevents us from exploring the possibility that its properties call for a more general equation of state such as the Van der Waals one. The aim of the present paper is to explore further this approach in order to see whether a cosmological fluid with Van der Waals state equation can be reconciled with observations. We have thus constrained the effective parameters of the theory looking for cosmological models which: $i)$ can admit a nowaday accelerating universe; $ii)$ satisfy the constraints on the estimated age of the Universe, $iii)$ are able to fit the data on the dimensionless coordinate distance to SNeIa and radio galaxies.

The paper is organized as follows. In Sect.\,2, we briefly summarize the main features of Van der Waals equation of state and choose the set of parameters which are best suited to assign the model. Dynamics of the model is described in Sect.\,3 where we determine the evolution of the energy density, the Hubble parameter and the scale factor. Matching with observations is performed in Sect.\,4 where the data on the dimensionless coordinate distance to SNeIa and radio galaxies are used to select among the models. In Sect.\,5, we compare the model prediction for the age of the universe with the estimated one, while, in Sect.\,6, we evaluate the position of the first three peaks of the CMBR anisotropy spectrum and qualitatively speculate on how structure formation could take place in the Van der Waals quintessence scenario. Sect.\,7 is then devoted to the discussion of the results and conclusions.

\section{The Van der Waals equation of state}

The dynamical system describing a Friedmann--Robertson--Walker (FRW) cosmology is given by the Friedmann equations \cite{CosmoBooks}\,:

\begin{equation}
\frac{\ddot{a}}{a} = - \frac{4 \pi G}{3} \ (\rho_b + \rho_{DM} + 3 p_{DM}) \ , 
\label{eq: fried1}
\end{equation}

\begin{equation}
H^2 + \frac{k}{a^2} = \frac{8 \pi G}{3} (\rho_b + \rho_{DM}) \ , 
\label{eq: fried2}
\end{equation}
and the continuity equations for each of the two fluids\,:

\begin{equation}
\dot{\rho_i} + 3 H \ (\rho_i + p_i) = 0 \ , 
\label{eq: continuity}
\end{equation}
where $H = \dot{a}/a$ is the Hubble parameter, the dot denotes the derivative with respect to cosmic time and $k = -1, 0 ,1$ is the spatial curvature constant respectively for open, flat and closed universes. Eqs.(\ref{eq: fried1}), (\ref{eq: fried2}) and (\ref{eq: continuity}) are derived by the Einstein field equations and the contracted Bianchi identities\footnote{Actually, the Bianchi identities lead to a conservation equation for the total energy density. The two substances are separately conserved only if we assume that the two fluids does not interact or interact very weakly. This is a quite reasonable assumption since popular dark matter candidates (such as neutrinos and WIMPs) are indeed very weakly interacting particles.} assuming that the source of the gravitational field is a a mixture of baryons with energy density $\rho_b$ and pressure $p_b = 0$ and dark matter with energy density $\rho_{DM}$ and pressure $p_{DM}$. To close the system and determine the evolution of the scale factor $a$ and of the other quantities of interest, the equation of state of the dark matter fluid (i.e. a relation between $\rho_{DM}$ and $p_{DM}$) is needed. 

In the standard cosmology, one assumes that the dark matter may be described as a perfect fluid so that the equation of state is $p_{DM} = \gamma \rho_{DM}$ where $0\leq \gamma\leq 1$ is the so called Zel'dovich interval. However, this is only an approximation which is not always valid and which does not describe the phase transitions between the successive thermodynamic state of the cosmic fluid. Several times, for instance at equivalence, two phases had to exist together. In these cases, a simple description by a perfect fluid equation of state is not realistic. An immediate generalization can be achieved by taking into account the Van der Waals equation of state which describes a two phase fluid. Also in this case, we have an approximation, but the consequences on dynamics are interesting \cite{CdMF,Kr03}. Hence we assume that the equation of state is\,:

\begin{equation}
p_{VdW} = \frac{\gamma \rho_{VdW}}{1 - \beta \rho_{VdW}} - \alpha \rho_{VdW}^2 \ ,
\label{eq: vdweq}
\end{equation}
which reduces to the perfect fluid case in the limit $\alpha,\beta \rightarrow 0$. Hereafter, we will denote with the subscript {\it b} ({\it VdW}) all the quantities referring to the baryons (the Van der Waals dark matter). In standard units, the $\alpha$ and $\beta$ coefficients may be rewritten as\,:

\begin{equation}
\alpha = 3 p_c \rho_c^{-2} \ \ ; \ \ \beta = (3 \rho_c)^{-1} \ ,
\label{eq: alphabeta}
\end{equation}
where $\rho_c$ and $p_c$ are the density and the pressure of the cosmic fluid at the Van der Waals critical point. The critical values are the indications that the cosmic fluid changes its phase at certain thermodynamic conditions. As a consequence, the three parameters $(\rho_c, p_c, \gamma)$ are not independent from each other. Inserting Eq.(\ref{eq: alphabeta}) into Eq.(\ref{eq: vdweq}) and considering the situation at the critical point, one gets\,:

\begin{equation}
p_c = \frac{3}{8} \gamma \rho_c \ , 
\label{eq: pcrhoc}
\end{equation}
so that the number of independent parameters is now reduced to two, which are $(\rho_c, \gamma)$. Using Eq.(\ref{eq: pcrhoc}) we may rewrite Eq.(\ref{eq: vdweq}) as\,:

\begin{equation}
\frac{p}{\rho_c} = \frac{3 \gamma \eta}{3 - \eta} - \frac{9}{8} \gamma \eta^2 
\label{eq: vdwour}
\end{equation}
having defined the dimensionless energy density $\eta \equiv \rho_{VdW }/\rho_c$. It is worth noting that\,:

\begin{displaymath}
p_{VdW} \sim \gamma \rho_{VdW} \ \ {\rm for} \ \ \eta << 1 \ \ , \ \ 
p_{VdW} \sim -\frac{9}{8} \gamma \rho_{VdW}^2  \ \ {\rm for} \ \ \eta >> 1 \ .
\end{displaymath}
As we know, observations tell us that we are living in a phase of accelerated expansion so that today the leading pressure must be negative. Let us first consider the case $\gamma > 0$ so that the pressure is positive (almost vanishing) in the limit $\rho_{VdW} << \rho_c$ and negative in the opposite limit $\rho_{VdW} >> \rho_c$. This simple consideration suggests us that models with $\gamma > 0$ could have some chance to reproduce the observed data provided that the present day energy density of the dark matter is much larger than the Van der Waals critical energy density, i.e. it should be $\eta_0 >> 1$ where hereon all quantities with a subscript 0 are evaluated today (i.e. at $z = 0)$. On the other hand, models with $\gamma < 0$ have a negative pressure in the limit $\eta << 1$ and a positive (and large) one in the limit $\eta >> 1$. We may thus conclude that, in order to give an accelerated expansion today, $\eta_0$ must be much smaller than 1 for models with $\gamma < 0$. It is worth noting that these conclusions have been obtained without the need to solve the Friedmann equations.

Data on the CMBR anisotropy spectrum strongly suggest that the universe is spatially flat today \cite{Boom,CMBR,WMAP,VSA} so that we will assume $k = 0$ in the rest of the paper. In this case, evaluating Eq.(\ref{eq: fried1}) at $z = 0$, we get\,:

\begin{equation}
\Omega_c \eta_0 + \Omega_{b,0} = 1 \ \rightarrow \Omega_c = \frac{1 - \Omega_{b,0}}{\eta_0}
\label{eq: ocob}
\end{equation}
with\,:

\begin{equation}
\Omega_c \equiv \rho_c/\rho_{crit}
\label{eq: etazero}
\end{equation}
with $\Omega_{b,0} = \rho_b(z = 0)/\rho_{crit}$ and $\rho_{crit} = 3 H_0^2/8 \pi G$ the critical density of the universe. It is worthwhile to note that Eq.(\ref{eq: ocob}) is a consequence of the assumption that there is only a single fluid other than baryons filling the spatially flat universe and playing the role of both dark matter and dark energy. From this point of view, Van der Waals quintessence may be considered in the framework of unified dark energy models such as the Chaplygin gas \cite{Chaplygin} and the Hobbit models \cite{Hobbit}. There is, however, a significant difference. The Van der Waals equation of state does not interpolate between a dust like pressure and a constant negative one as $\rho_{VdW}$ evolves. Indeed, for models with $\gamma > 0$ the pressure becomes quite small for $\eta << 1$, but it is not zero. On the other hand, when $\gamma < 0$, the pressure becomes quite small and negative, but still remains different from zero.

It is worth noting that it is possible to start narrowing the class of models to explore without the need to explicitly solve the Friedmann equations. To this aim, let us consider the deceleration parameter\,:

\begin{equation}
q \equiv - \frac{a \ddot{a}}{\dot{a}^2} = - \frac{\ddot{a}}{a} \ \frac{1}{H^2} =\frac{1}{2} + \frac{3}{2} \frac{p_{VdW}}{\rho_b + \rho_{VdW}}
\label{eq: defq}
\end{equation}
where we have combined Eqs.(\ref{eq: fried1}) and (\ref{eq: fried2}). Inserting Eq.(\ref{eq: vdwour}) into Eq.(\ref{eq: defq}), evaluating the result at the present day and solving with respect to $\gamma$, we get\,:

\begin{equation}
\gamma = \frac{8 (1 - 2 q_0) (\eta_0 - 3)}{9 (1 - \Omega_{b,0}) (3 \eta_0^2 - 9 \eta_0 + 8)} \ .
\label{eq: gammavsqz}
\end{equation}
Since $\Omega_{b,0} < 1$, the denominator is always positive so that $\gamma$ is consistently defined by Eq.(\ref{eq: gammavsqz}). We may thus use $q_0$ as a parameter instead of $\gamma$ and characterize the models by the three parameters $(q_0, \log{\eta_0}, \Omega_{b,0})$ where we use the logarithm of $\eta_0$ instead of $\eta_0$ itself for reasons that will be clear later. It is worth noting that using $q_0$ instead of $\gamma$ allows to select immediately nowadays accelerating models. Moreover, it is easier to choose a range for $q_0$ than for $\gamma$ that is {\it a priori} completely unknown. Actually, there is a third parameter that has to be given to completely assign the model and study its dynamics. This is the Hubble constant $H_0$ which, however, we will consider as a known quantity. We will explain later why this assumption is necessary and why, nonetheless, it does not introduce any loss of generality. 

Finally, let us observe that there is another interesting quantity that may be evaluated without the need of solving the Friedmann equations. By definition, the barotropic factor\footnote{We drop the subscript {\it VdW} because there is no possibility of confusion with the barotropic factor of the baryons that is identically zero.} $w(z)$ of the Van der Waals fluid is given as\,:

\begin{displaymath}
w(z) \equiv \frac{p_{VdW}}{\rho_{VdW}} = \frac{3 \gamma}{3 - \eta(z)} - \frac{9}{8} \gamma \eta(z) \ .
\end{displaymath} 
Using Eq.(\ref{eq: gammavsqz}) and evaluating the result at the present day, we get\,:

\begin{equation}
w_0 = - \frac{1 - 2 q_0}{3 (1 - \Omega_{b,0})} \ .
\label{eq: wzqz}
\end{equation}
For accelerating models ($q_0 < 0$), $w_0$ takes on negative values and may also be lower than -1 thus suggesting that phantom models (that are indeed characterized by negative pressure fluids with $w_0 < -1$) could be somewhat confused with Van der Waals quintessence.

\begin{figure}[!t]
\begin{center}
\includegraphics{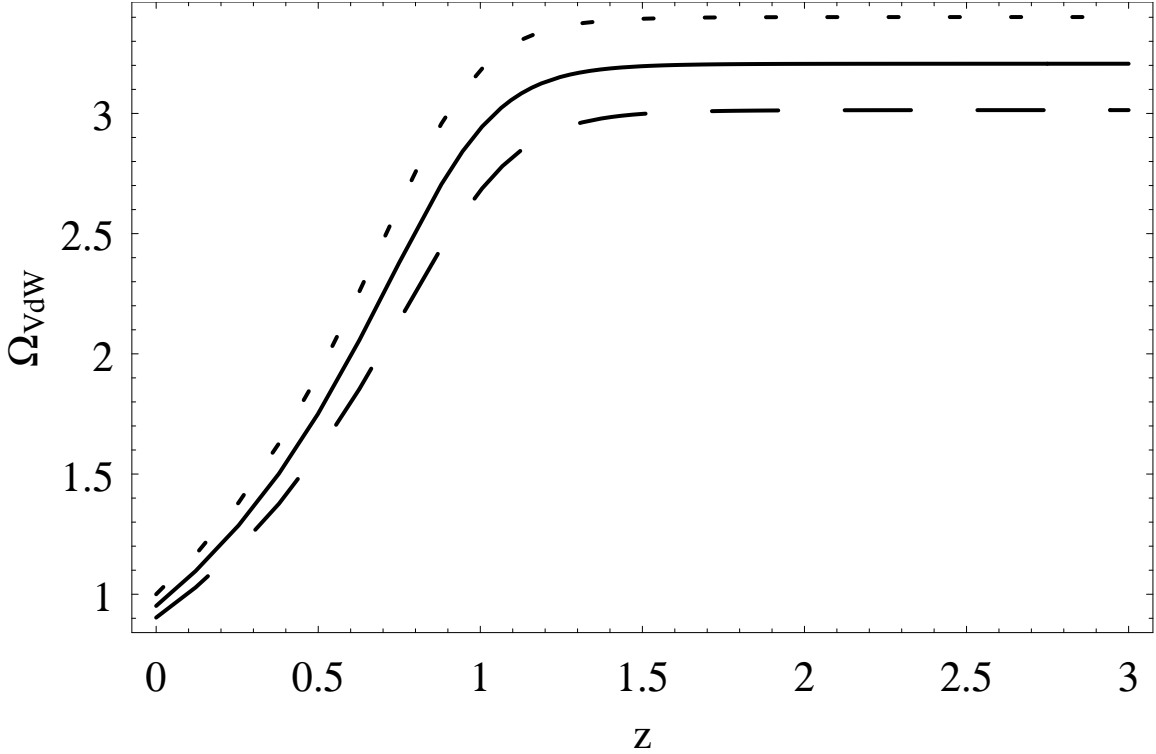}
\end{center}
\caption{The evolution against the redshift of the Van der Waals density parameter setting $(q_0, \log{\eta_0}) = (-0.38, -0.11)$ and three different values of $\Omega_{b,0} h^2$, namely $\Omega_{b,0} h^2 = 0$ (short dashed), $\Omega_{b,0} h^2 = 0.0214$ (solid) and $\Omega_{b,0} h^2 = 0.0428$ (long dashed). We use $h = 0.664$ to get $\Omega_{b,0}$ from $\Omega_{b,0} h^2$.}
\label{fig: ez}
\end{figure}

\section{The dynamics of the universe}

Having described the general features of the model, let us now determine the dynamics of the universe, i.e. let us investigate how the relevant physical quantities (scale factor, energy densities and Hubble parameter) evolve. As a preliminary step, let us remember that, solving the continuity equations for baryons, we get\,:

\begin{equation}
\Omega_b(z) = \Omega_{b,0} (1 + z)^3
\label{eq: obvsz}
\end{equation} 
with $z = 1/a - 1$ the redshift. To solve the continuity equation for the Van der Waals dark matter, it is convenient to change variable from the cosmic time $t$ to the redshift $z$ and to use the dimensionless energy density $\eta(z)$. We thus get\,:

\begin{equation}
\frac{d\eta}{dz} = \frac{3 \eta \left [ 9 \gamma \eta^2 - (27 \gamma + 8) \eta + 24 (1 + \gamma) \right ]}{8 (1 + z) (3 - \eta)} \ .
\label{eq: etavsz}
\end{equation}
This is a first order nonlinear differential equation for $\eta$ which may be numerically solved provided that the three parameters $(q_0, \log{\eta_0}, \Omega_{b,0})$ are given. It is worth stressing that, although not explicitly present, the parameter $\Omega_{b,0}$ enters Eq.(\ref{eq: etavsz}) through $\gamma$. Moreover, being Eq.(\ref{eq: etavsz}) nonlinear, a sort of {\it butterfly effect} takes place with the result that also small changes in the baryons content lead to significantly different evolutions of the the energy density of the Van der Waals dark matter with the redshift $z$. This is clearly shown in Fig.\,\ref{fig: ez} where we plot $\Omega_{VdW}(z) = \Omega_c \eta(z)$ for a particular choice of $(q_0, \log{\eta_0})$ and three different values of $\Omega_{b,0} h^2$. Comparing the short dashed line (corresponding to a model with no baryons) with the other two, we may safely conclude that, although the present day baryon density parameter is quite small ($\Omega_{b,0} \sim 0.04$), it can not be neglected without introducing a severe bias in the past ($z > 1$) evolution of the Van der Waals energy density\footnote{Note that we are defining $\Omega_{VdW}(z)$ as the ratio between the Van der Waals energy density and the present day critical density $\rho_{crit} = 3 H_0^2/8 \pi G$. It is thus not surprising that $\Omega_{VdW}(z)$ gets larger than 1 even if the universe is assumed to be spatially flat.}. This is also evident considering the the Hubble parameter. While, for $z < 1$, there is essentially no difference at all among models having the same values of $(q_0, \log{\eta_0})$ but different baryon content, the dependence on $\Omega_{b,0}$ is more and more important going back in time, i.e. to higher redshifts. In particular, we note that, for a given $z$, the higher is $\Omega_{b,0}$, the larger is $H(z)$, i.e. models with higher baryons content evolve faster. It is worth noting that these results do not depend on the values of the parameters $(q_0, \log{\eta_0})$ that only determine when the effect of $\Omega_{b,0}$ becomes important. 

\begin{figure}[!t]
\begin{center}
\includegraphics{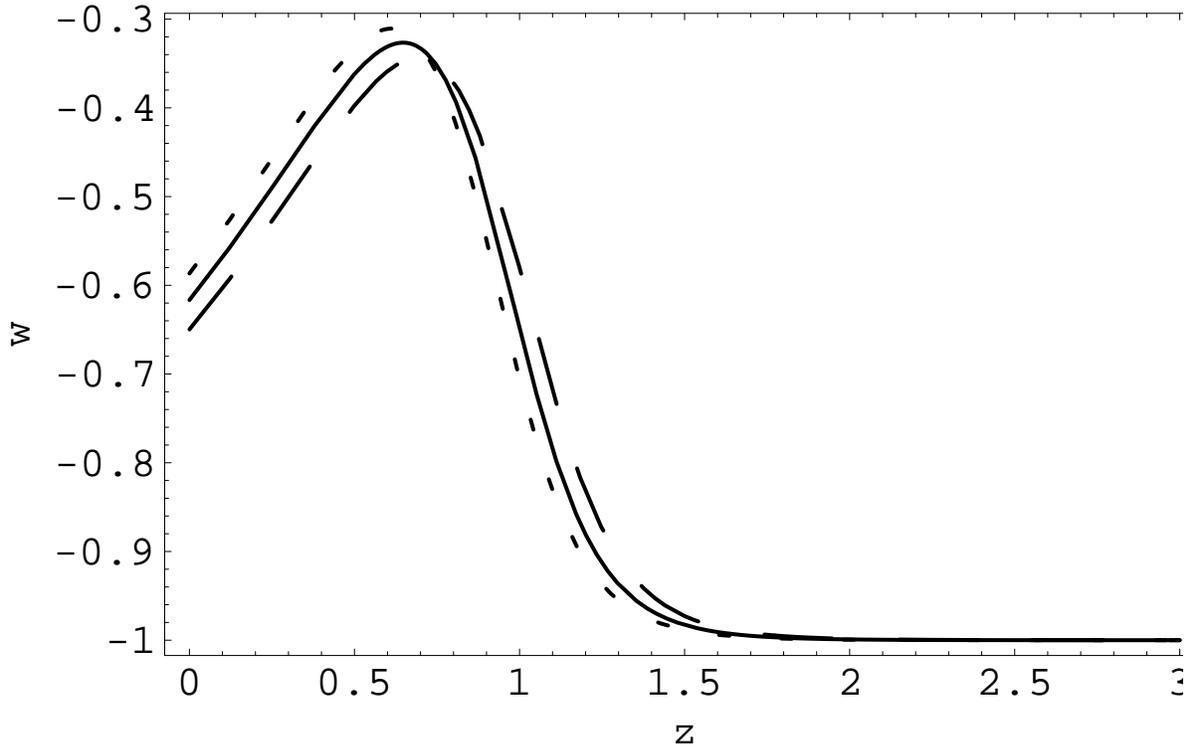}
\end{center}
\caption{The evolution against the redshift of the barotropic factor $w$ for the same models considered in Fig.\,\ref{fig: ez}.}
\label{fig: wz}
\end{figure}

A striking feature of Fig.\,\ref{fig: ez} is the flattening of $\Omega_{VdW}(z)$ towards a constant value suggesting that the Van der Waals fluid mimics a cosmological constant term in the past. {\bf It is worth stressing that this behaviour does not depend on the particular choice of the parameters $(q_0, \log{\eta_0})$, but is rahter a characteristic feature of the Van der Waals fluid. However, the lower is the value of $\log{\eta_0}$, the higher is the redshift $z$ when $\Omega_{VdW}(z)$ becomes constant. Although quite peculiar, this result may be easily explained considering the evolution of the barotropic factor $w(z)$. As an example, Fig.\,\ref{fig: wz} shows $w(z)$ for the same models considered in Fig.\,\ref{fig: ez}. First, we note that the effect of $\Omega_{b,0}$ is negligible (for realistic values of this quantity) in agreement with what is predicted for $w_0$ by Eq.(\ref{eq: wzqz}). Moreover, $w(z)$ quickly converges to $w_{\Lambda} = -1$ so that the Van der Waals fluid behaves as the cosmological constant in the past. Changing the values of the parameters $(q_0, \log{\eta_0})$ have the only effect of increasing or decreasing the rate of convergence, but not the asymptotic limit of $w(z)$. In particular, implicitly defining $z_{\Lambda}$ as $w(z_{\Lambda}) = -1$, it is possible to show that $z_{\Lambda}$ is a decreasing function of $\log{\eta_0}$, while it depends only weakly on $q_0$. Since $\Omega_{VdW}(z)$ becomes constant for $z > z_{\Lambda}$, this explains why the energy density of the Van der Waals fluid becomes flat in Fig.\,\ref{fig: ez} for higher values of $z$ as $\log{\eta_0}$ decreases. We may thus conclude that the Van der Waals fluid behaves as the cosmological constant at high redshift whatever are the values of the parameters $(q_0, \log{\eta_0}, \Omega_{b,0})$.} This important result will have profound impact on the evaluation of the positions of the peaks of the CMBR anisotropy spectrum as we will discuss later in Sect.\,6.

\begin{figure}[!t]
\begin{center}
\includegraphics{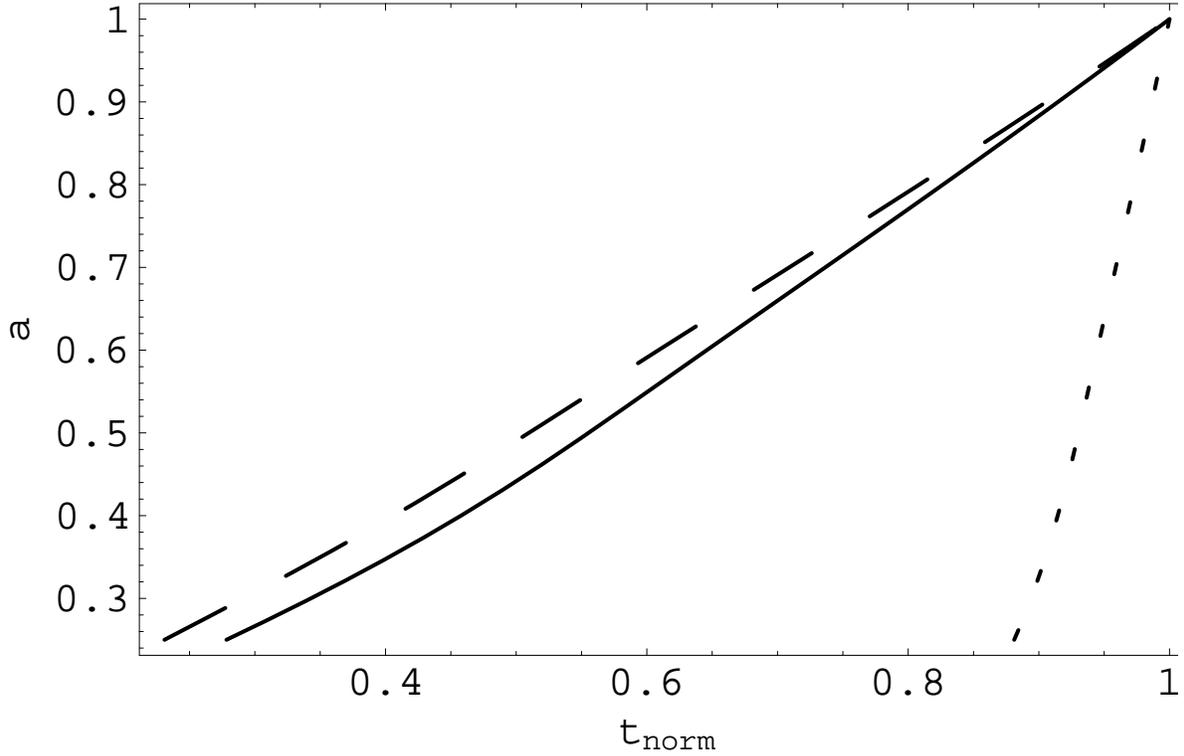}
\end{center}
\caption{The evolution of the scale factor $a$ with the scaled cosmic time $t_{norm} = t/t_0$ for the same models considered in Fig.\,\ref{fig: ez}.}
\label{fig: at}
\end{figure}

Having determined $H(z)$, it is straightforward to estimate the age of the universe at redshift $z$ as\,:

\begin{equation}
t(z) = \int_{z}^{\infty}{\frac{1}{(1 + \zeta) H(\zeta)} d\zeta} = t_H \int_{z}^{\infty}{\frac{1}{(1 + \zeta) E(\zeta)} d\zeta}
\label{eq: age}
\end{equation} 
where $t_H = c/H_0 = 9.78 h^{-1} \ {\rm Gyr}$ is the Hubble time (being $h$ the Hubble constant in units of $100 \ {\rm km \ s^{-1} \ Mpc^{-1}}$) and $E(z) \equiv H(z)/H_0$. The present age of the universe is then $t_0 = t(z = 0)$. The evolution of the scale factor $a$ as function of cosmic time $t$ may be obtained by numerically inverting Eq.(\ref{eq: age}) and remembering that $a = (1 + z)^{-1}$. The result is shown in Fig.\,\ref{fig: at} for the same models considered above. It is worth noting that $a(t)$ does not enter anyone of the tests we will perform later that are only dependent on $H(z)$. Nonetheless, it is interesting to look at the scale factor to get a feeling of how the universe evolves with time. 

It turns out that neglecting baryons leads to a strong error in the determination of the scale factor and thus on the age of the universe. The {\it butterfly effect} here works quite hard. Actually, it is worth noting that the shape of $a(t)$ does not depend strongly on the exact value of $\Omega_{b,0}$ as could be inferred comparing the solid and long dashed lines in Fig.\,\ref{fig: at}. Doubling $\Omega_{b,0}$ does not change significantly $a(t/t_0)$ so that we may conclude that what is important is to take into account the presence of baryons, but not the precise value of their density parameter (unless we take completely unrealistic values).

\section{The dimensionless coordinate distance}

A whatever model that aims at describing the evolution of the universe must be able to reproduce what is indeed observed. It is thus mandatory to test the viability of the proposed Van der Waals quintessence by contrasting and comparing it to the astrophysical data available up to now. This is also a powerful tool to constrain the model parameters thus paving the way towards a complete characterization of the model. As a first step, we may resort to the Hubble diagram of SNeIa, that is the plot of the distance modulus as function of the redshift $z$. However, we prefer here to follow a very similar, but more general approach considering as cosmological observable the dimensionless coordinate distance defined as\,:

\begin{equation}
y(z) = \int_{0}^{z}{\frac{1}{E(\zeta)} d\zeta} \ .
\label{eq: defy}
\end{equation}
with $E(z)$ defined above. For completeness, we remember that $y$ is related to the usual luminosity distance $D_L$ (which is the quantity measured through the SNeIa distance modulus) as follows\,:

\begin{displaymath}
D_L = \frac{c}{H_0} (1 + z) y(z) \ .
\end{displaymath}
It is worth noting that $y(z)$ does not depend explicitly on $H_0$ so that it is now clear why we are confident that our above choice for $H_0$ does not alter the main result. Actually, $H_0$ enters in the estimate of $y_{obs}(z_i)$, the observed dimensionless coordinate distance to an object at redshift $z_i$. Daly \& Djorgovski \cite{DD04} have determined $y(z)$ for the SNeIa in the Gold dataset of Riess et al. \cite{Riess04} which represents the most updated and homogeneous SNeIa sample today available. Since SNeIa allows to estimate $D_L$ rather than $y$, a value of $H_0$ has to be set. Fitting the Hubble law to a large set of low redshift ($z < 0.1$) SNeIa, Daly \& Djorgovski \cite{DD04} have determined\,:

\begin{displaymath}
H_0 = 66.4 \pm 0.8 \ {\rm km \ s^{-1} \ Mpc^{-1}} \ . 
\end{displaymath}
We thus set $h = 0.664$ in order to be consistent with Daly \& Djorgovski \cite{DD04}, but we have checked that varying $h$ in the $68\%$ CL quoted above does not alter the main results\footnote{Note that the value we are using is consistent also with $H_0 = 72 \pm 8 \ {\rm km \ s^{-1} \ Mpc^{-1}}$ given by the HST Key project \cite{Freedman} based on the local distance ladder.}.

To increase the sample, Daly \& Djorgovski added 20 further points on the $y(z)$ diagram using a technique based on the angular dimension of radiogalaxies \cite{DD04,RGdata}. Both this sample and the 157 SNeIa contained in the Riess et al. compilation span the redshift range $(0.1, 1.8)$ so that it is possible to detect eventual systematic deviations of one tracer from another. None of such trends have been detected so that the full sample may be used without introducing spurious problematic features in the $y(z)$ diagram. 

To determine the best fit parameters, we define the following merit function\,:

\begin{equation}
\chi^2(q_0, \log{\eta_0}) = \frac{1}{N - 2} \sum_{i = 1}^{N}{\left [ \frac{y(z_i; q_0, \log{\eta_0}) - y_i}{\sigma_i} \right ]^2}
\label{eq: defchi}
\end{equation} 
where the observed quantities $(z_i, y_i, \sigma_i)$ are given in Tables\,1 and 2 of Ref.\,\cite{DD04}. Note that, although similar to the usal reduced $\chi^2$ introduced in statistics, the $\chi^2$ defined above is not forced to be 1 for the best fit model since the uncertainties $\sigma_i$ are not Gaussian distributed, but take care of both statistical errors and systematic uncertainties. Nonetheless, it is possible to compare different couples of model parameters on the base of the $\chi^2$ value. To determine constraints on the model parameters, we first define the marginalized likelihood functions\,:

\begin{eqnarray}
{\cal{L}}_q(q_0) \propto \int{\exp{\left [ -\chi^2(q_0, \log{\eta_0})/2 \right ]} \ d\log{\eta_0}} \ , \nonumber \\ 
~ \\
{\cal{L}}_{\eta}(\log{\eta_0}) \propto \int{\exp{\left [ -\chi^2(q_0, \log{\eta_0})/2 \right ]} \ dq_0} \ . \nonumber 
\label{eq: deflike}
\end{eqnarray}
After having normalized to 1 at maximum, the $68\%$ CL ($95\%$ CL) on a parameter $p_i$ is obtained by solving for ${\cal{L}}_i(p_i) = \exp{(-0.5)}$ (${\cal{L}}_i(p_i) = \exp{(-2)}$). 

Having set the Hubble constant as above, we are left with a three dimensional space to explore defined by the model parameters $(q_0, \log{\eta_0}, \Omega_{b,0})$. Actually, we have seen that the main dynamical quantities we are interested in depend only weakly on $\Omega_{b,0}$ in the redshift range that is proven by the available astrophysical data. Moreover, $\Omega_{b,0}$ is severely constrained by theoretical models of nucleosynthesis and by the observed abundance of light elements. Based on these considerations, Kirkman et al. \cite{Kirk} have estimated\,:

\begin{displaymath}
\Omega_{b,0} h^2 = 0.0214 \pm 0.0020 \ .
\end{displaymath}
Neglecting the small error, we thus set $\Omega_{b,0} h^2 = 0.0214$ and use the above quoted value of $h$ to get our estimate of $\Omega_{b,0}$. We have therefore only two parameters to constrain, namely $q_0$ and $\log{\eta_0}$. Note that we use $\log{\eta_0}$ instead of $\eta_0$ itself since the former is easier to handle in numerical codes. As regard the range for $q_0$, since we are interested in accelerating models only, we set $q_0 = 0$ as upper limit. A lower limit may be obtained by the following argument. Let us insert Eq.(\ref{eq: gammavsqz}) into Eq.(\ref{eq: etavsz}) and evaluate the result at $z = 0$. We get\,:

\begin{displaymath}
\left ( \frac{d\eta}{dz} \right )_{z = 0} = \frac{2(1 + q_0) - 3 \Omega_{b,0}}{1 - \Omega_{b,0}} \times \eta_0 \ .
\end{displaymath} 
It is reasonable to impose that the energy density is a decreasing function of cosmic time so that it is always $d\eta/dz > 0$. In order to fulfill this condition at $z = 0$, we must impose $q_0 > (3 \Omega_{b,0} - 2)/2 \simeq -1$. Eq.(\ref{eq: wzqz}) makes it possible to see that the above lower limit on $q_0$ excludes all models with $w_0 < -1$. While theoretically motivated by the need of not violating the energy conditions, such a choice seems to be disfavoured by the observations. Actually, fitting models with constant equation of state to the SNeIa Hubble diagram and the CMBR anisotropy spectrum gives $w_0 = -1.43_{-0.38}^{+0.16}$ \cite{HM04}. It is worth noting, however, that this estimate is strongly model dependent so that it does by no means imply that $w_0$ should be in that range also for the Van der Waals quintessence scenario we are investigating. Therefore, we have decided to use the above quoted lower limit for $q_0$ although this excludes models with $w_0 < -1$. Nonetheless, we have checked that the constraints on $q_0$ from observations are unaltered if we decrease the lower limit up to $q_0 = -2$ thus entering the realm of $w_0 < -1$ models.

\begin{figure}[!t]
\begin{center}
\includegraphics{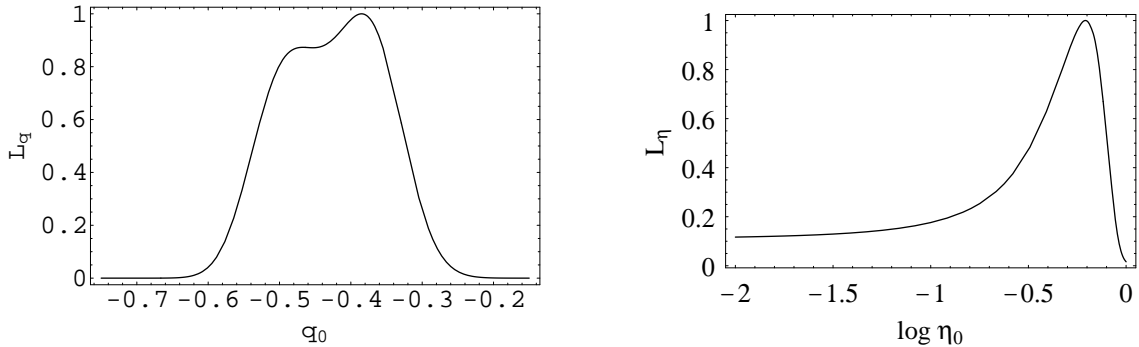}
\end{center}
\caption{Marginalized likelihood functions (normalized at 1 at maximum) for the two model parameters $q_0$ (left panel) and $\log{\eta_0}$ (right panel).}
\label{fig: like}
\end{figure}

There are no physical motivations to select a plausible range for $\log{\eta_0}$ so that we have resorted to a trial and error procedure trying to not exclude any interesting regions. We thus end up with the following range for the model parameters\,:

\begin{equation}
(3 \Omega_{b,0} - 2)/2 \le q_0 \le 0 \ \ \ \ , \ \ \ \ -3.0 \le \log{\eta_0} \le 0.0 \ .
\label{eq: rangetot}
\end{equation}
By running $(q_0, \log{\eta_0})$ in the range defined above, we find the best fit parameters\,:

\begin{figure}[!t]
\begin{center}
\includegraphics{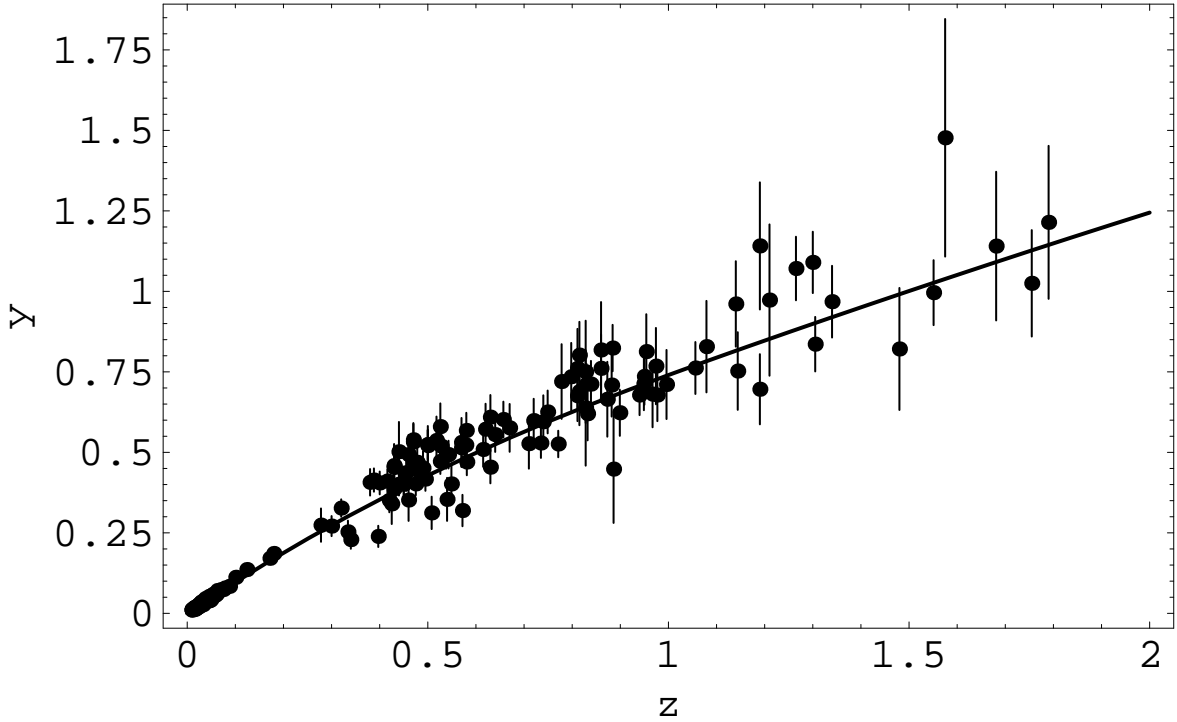}
\end{center}
\caption{Observed data and theoretical curve of the dimensionless coordinate distance for the best fit Van der Waals model defined by $(q_0, \log{\eta_0}) = (-0.38, -0.11)$ with $(h, \Omega_{b,0}) = (0.664, 0.0214)$.}
\label{fig: bestfit}
\end{figure}

\begin{displaymath}
(q_0, \log{\eta_0}) = (-0.38, -0.21) 
\end{displaymath}
giving $\chi^2 = 1.26$. The fit is quite successful as can be seen in Fig.\,\ref{fig: bestfit}, while Fig.\,\ref{fig: like} shows the normalized likelihood functions from which we derive\,:

\begin{eqnarray}
-0.52 \le q_0 \le -0.33 \ \ , \ \ -0.42 \le \log{\eta_0} \le -0.11 \ \ & \ \ {\rm at \ 68\% \ CL} \ \ , \nonumber \\ 
~ \\
-0.58 \le q_0 \le -0.28 \ \ , \ \ -1.38 \le \log{\eta_0} \le -0.05 \ \ & \ \ {\rm at \ 95\% \ CL} \ \ . \nonumber
\label{constraints}
\end{eqnarray}
Note that the 95$\%$ upper and lower limits on $q_0$ are significantly far away from those of the range (\ref{eq: rangetot}) which is a reassuring evidence suggesting that we have not excluded any observationally interesting region of the parameter space. While this is also true for the 95$\%$ lower limit on $\log{\eta_0}$, the corresponding upper limit is quite near to the upper edge in Eq.(\ref{eq: rangetot}) which could be somewhat worrisome. However, the shape of the likelihood function for $\log{\eta_0}$ suggests that this is not a problem since the excluded region is outside the 99$\%$ CL. As a further consistency check, we have also repeated the fit adopting an higher upper limit on $\log{\eta_0}$, but the constraints (20) turn out to be unaltered.

The first remarkable result of the fit is the exclusion of models with positive values of $\log{\eta_0}$, i.e. models having a nowaday energy density of the Van der Waals fluid larger than the Van der Waals critical density $(\eta_0 > 1)$ are not compatible with the observed data. Moreover, the test is not very efficient to constrain the model parameters since only a minor part of the range for $q_0$ is cut away, while $\log{\eta_0}$ is allowed to vary over almost an order of magnitude. Actually, this is not surprising. In the redshift range probed by our data ($z < 1.8$), the dimensionless coordinate distance for two models with very different values of $\log{\eta_0}$ approximately coincide within the observational errors so that, to discriminate among distinct models, we need either more precise data in the redshift range $(1.0, 2.0)$ or extending the dataset to $z$ up to $\sim 3$. While it is likely that forthcoming satellite experiments such as the planned SNAP mission \cite{SNAP} will furnish more precise measurements of $y$ in the redshift range $(1.0, 2.0)$, going to higher redshift will need a tracer other than SNeIa. Good candidates in this sense are compact radio sources \cite{AngSizeTest} since they can be detected up to $z \sim 4$, but there are still some problems related to the evolution with redshift of their physical properties. 

\section{The age of the universe}

The present day age of the universe $t_0$ is a powerful tool to constrain the parameters of a given cosmological model. Let us thus investigate what can be learned by applying this test to the Van der Waals scenario we are considering.

First, we need an estimate of $t_0$ from observational data. In Ref.\,\cite{VSA}, Rebolo et al. have performed a detailed combined analysis of the WMAP and VSA data on the CMBR anisotropy spectrum and SDSS galaxy clustering thus obtaining $t_0 = 14.4^{+1.4}_{-1.3} \ {\rm Gyr}$ at $68\%$ confidence limit. A more precise determination has been obtained by Seljak et al. \cite{Sel04} who have fitted the $\Lambda$CDM model to a combined dataset comprising the  CMBR anisotropy spectrum, the galaxy power spectrum, the SNeIa Hubble diagram, the dependence of the galaxy bias on mass and the Ly$\alpha$ clouds power spectrum. As a result, they get $t_0 = 13.6 \pm 0.19 \ {\rm Gyr}$ at $68\%$ CL. Actually, both these estimates are model dependent since the authors assume {\it a priori} a background cosmological model and then determine $t_0$ from the best fit parameters of that model. In order to avoid any systematic bias, we prefer to use a model independent estimate although this leads to enlarge the error bars. To this end, we resort to age estimates of globular clusters and, following Krauss \cite{Krauss}, we retain as viable all models such that\,:

\begin{figure}[!t]
\begin{center}
\includegraphics{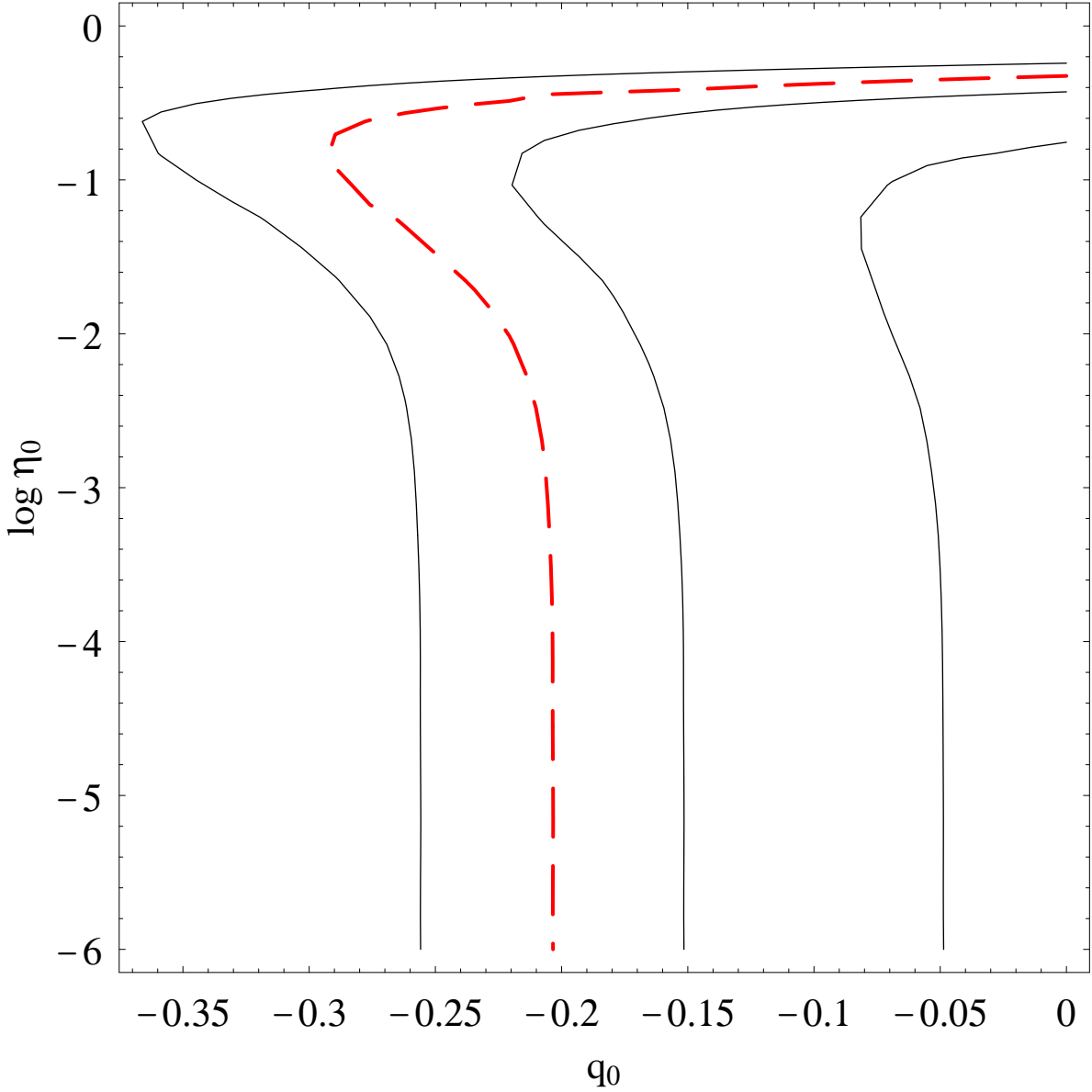}
\end{center}
\caption{Contour plots of equal $t_0$ in the plane $(q_0, \log{\eta_0})$. The solid lines refer to values from 14.5 to 16.5 Gyr (in steps of 1 Gyr) from right to left. The dashed line is the observed upper limit $t_0 = 16.0 \ {\rm Gyr}$.}
\label{fig: agefig}
\end{figure}

\begin{equation}
t_0 \in (10.2, 16.0) \ {\rm Gyr} \ .
\label{ageconstr}
\end{equation}
Note that the range quoted above is consistent both with the model dependent estimates of Rebolo et al. and Seljak et al. and with $t_{0} > 12.5 \pm 3.5 \ {\rm Gyr}$ from radioisotopes studies \cite{Cayrel} so that we are confident not to be excluding interesting models.

Given $(q_0, \log{\eta_0})$, the predicted age of the universe for the corresponding Van der Waals model may be estimated setting $z = 0$ in Eq.(\ref{eq: age}). Fig.\,\ref{fig: agefig} shows the age contours in the $(q_0, \log{\eta_0})$ plane for the model with $(h, \Omega_{b,0})$ set as discussed above. Note that we have greatly enlarged the range for $\log{\eta_0}$ in order to explore with much detail this poorly constrained parameter. The results of the age test are somewhat surprising. First, positive values of $\log{\eta_0}$ are strongly excluded since they give rise to values of $t_0$ of the order of hundreds of Gyr in striking disagreement with Eq.(\ref{ageconstr}). Furthermore, the age test also cuts away a large part of the range for $q_0$ accepting only models with $q_0 > -0.29$. Roughly, we can summarize the results of this test giving the following constraints\,:

\begin{equation}
-0.29 \le q_0 \le 0 \ \ , \ \ -6.0 \le \log{\eta_0} \le -0.5 \ .
\label{eq: rangetime}
\end{equation}
Note that this is actually a little bit larger than the region delimited by the dashed line in Fig.\,\ref{fig: agefig} since we have approximated it as a rectangular one, while it is not. However, this simplification will not alter our main results. Note that, for the model parameters in the range (\ref{eq: rangetime}), $p_{VdW,0} \simeq \gamma \rho_{VdW,0}$ with $\gamma < 0$ so that the Van der Waals fluid behaves essentially as a perfect fluid with negative pressure. 

{\bf Comparing Eq.(\ref{eq: rangetime}) with the constraints (20) from the fit to the dimensionless coordinate distance shows that the two ranges have a good overlap in $\log{\eta_0}$ at 95$\%$ CL, but the constraints on $q_0$ are marginally consistent only at the 95$\%$ CL.} Moreover, all the allowed models predict values of $t_0$ close the upper end in Eq.(\ref{ageconstr}). One should thus conclude that, in order to reconcile Van der Waals quintessence with both the SNeIa and radiogalaxies distance data and the age of the universe, a certain degree of fine tuning of the parameters is needed. Although this is not a problem in its own, it is worth noting that the Van der Waals equation of state is only an approximation to the actual unknown equation of state of the dark matter component which is likely to be valid only over a limited range of the evolution history of the universe. Unfortunately, a detailed knowledge of the dark matter thermodynamical properties and of their possible dependence on $z$ is completely lacking so that it is impossible to say what is the redshift range over which Eq.(\ref{eq: vdwour}) holds. This systematic error does not affect the fit to the dimensionless coordinate distance since this test only probes a limited and nearby redshift range, but could have a profound impact on the estimate of $t_0$ since this latter depends on the full evolutionary history. As a consequence, we do not consider as a serious shortcoming the only marginal agreement of predicted and estimated age of the universe for the Van der Waals models selected by the fit to the SNeIa and radiogalaxies dimensionless coordinate distances data.

\section{The peaks of the CMBR and some remarks on structure formation}

The cosmological model we have discussed insofar is made out of matter only, but we have made a clear separation between the baryons and dark matter. While the former is still described as dust matter, the equation of state of the latter is the Van der Waals one so that the properties of the dark matter in this scenario are radically different from that of the standard cold dark matter (CDM). Indeed, while for CDM it is $w = 0$ at all $z$, for the Van der Waals dark matter $w$ depends on $z$ and quickly approaches the value $w = -1$ as shown in Fig.\,\ref{fig: wz}. It thus makes sense to ask how structure formation evolves in the scheme we are proposing. This is a quite complicated task and will be addressed in detail in a forthcoming paper. Nonetheless, here we give some qualitative comments to illustrate the subtleties of this topic.

\begin{figure}[!t]
\begin{center}
\includegraphics{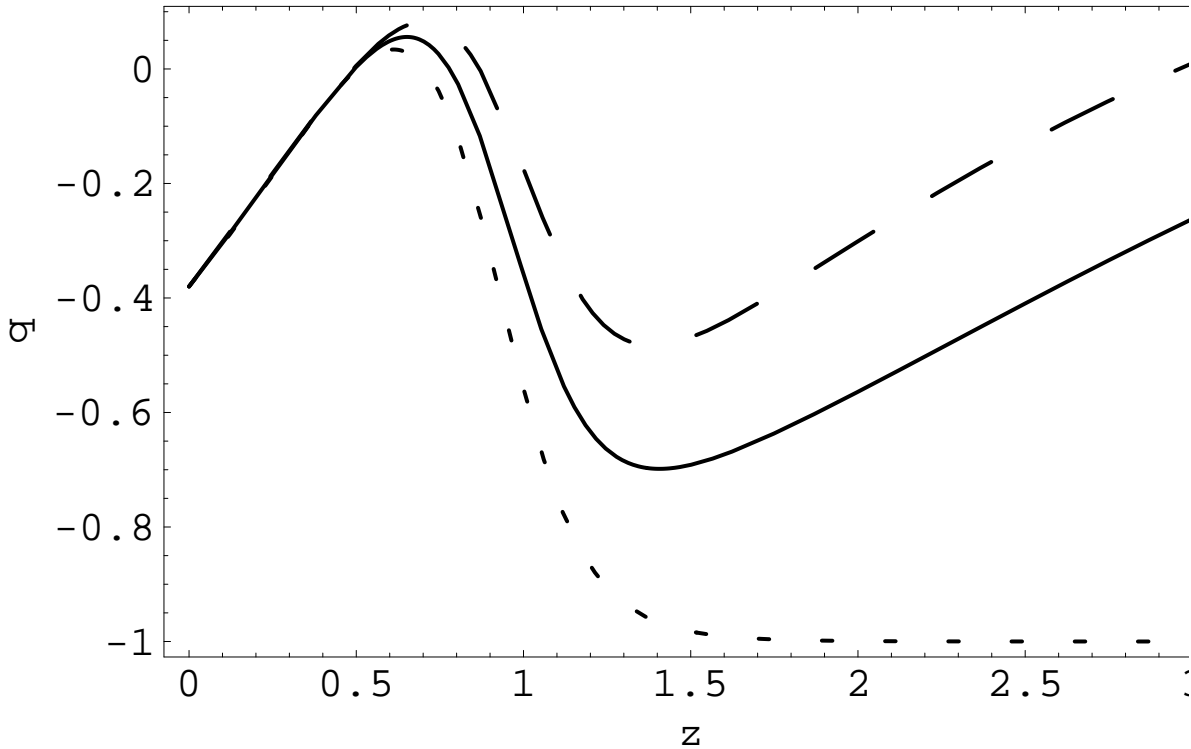}
\end{center}
\caption{Deceleration parameter $q$ as function of the redshift $z$ for the best fit model with $(q_0, \log{\eta_0}) = (-0.38, -0.21)$ and three different values of $\Omega_{b,0} h^2$, namely $\Omega_{b,0} h^2 = 0$ (short dashed), $\Omega_{b,0} h^2 = 0.0214$ (solid) and $\Omega_{b,0} h^2 = 0.0428$ (long dashed). We set $h = 0.664$ to convert from $\Omega_{b,0} h^2$ to $\Omega_{b,0}$.}
\label{fig: decfig}
\end{figure}

As a first remark, let us remember that structure formation may efficiently take place only during a decelerating phase of the universe evolution. Moreover, the SNeIa Hubble diagram shows some evidences of a transition from acceleration to deceleration although the estimate of the transition redshift $z_T$ (defined so that $q(z_T) = 0$) are quite model dependent. For instance, Riess et al. \cite{Riess04} obtain $z_T = 0.46 \pm 0.13$ by using the ansatz $q(z) = q_0 + (dq/dz)_{z = 0} z$, while the detailed analysis of the $\Lambda$CDM model performed by Seljak et al. \cite{Sel04} gives $0.52 \le z_T \le 0.91$. In Fig.\,\ref{fig: decfig}, we report the deceleration parameter for the best fit model parameters $(q_0, \log{\eta_0})$ and three different baryon contents. For this particular model, $q(z)$ takes positive  values only in the redshift range $(\sim 0.4, \sim 0.7)$ so that the universe is accelerating also during the period when structure formation takes place in the concordance $\Lambda$CDM model. The situation is partially ameliorated considering models with larger values of $\log{\eta_0}$ since the redshift range over which $q(z)$ is positive enlarges and shifts to higher $z$. However, one should also take into account the effect of $\Omega_{b,0} h^2$ since this could significantly impact the behaviour of the deceleration parameter. 

Likely, the most worrisome feature of the Van der Waals quintessence scenario is the barotropic factor $w$ going to -1 in the past since this makes the Van der Waals fluid never behaving as standard cold dark matter which could suggest a completely unrealistic growth of perturbations in our scenario. However, such a conclusion is at least premature. Let us consider the $\Lambda$CDM model that correctly describes the evolution of structures as we observe it. There are two fluids, standard cold dark matter with $p = 0$ and the cosmological constant with $p = -\rho$, and during the structure formation epoch the CDM energy density dominates over that of the cosmological constant. Actually, a very similar situation takes place in our model. Indeed, we have checked that, in the redshift range where presumably structure formation takes place, the baryons energy density dominates over that of the Van der Waals fluid that, in this period, is very well approximated by a cosmological constant\,-\,like term for all values of the model parameters. Therefore, in the far past our model is formally equivalent to the $\Lambda$CDM model with the baryons and the Van der Waals dark matter playing the roles of CDM and $\Lambda$ respectively. This nice result suggests that structure formation could evolve in a very similar way, but a detailed investigation is needed to draw a definitive answer. 

As a first preliminary investigation, we may compute the positions of the first three peaks in the CMBR anisotropy spectrum using the procedure detailed in Refs.\,\cite{DorLil01,DorLil02}. According to this prescription, in a flat universe made out of a matter term and a scalar field\,-\,like fluid, the position of the $m$\,-\,th peak is given by\,:

\begin{equation}
l_m = l_A (m - \bar{\varphi} - \delta \varphi_m)
\label{eq: lm}
\end{equation}
with $l_A$ the acoustic scale, $\bar{\varphi}$ the overall peak shift and $\delta \varphi_m$ the relative shift of the $m$\,-\,th peak with respect to the first. While $\bar{\varphi}$ and $\delta \varphi_m$ are given by the approximated formulae in Ref.\,\cite{DorLil02}, the acoustic scale for flat universes may be evaluated as \cite{DorLil01}\,:

\begin{equation}
l_A = \pi \bar{c}_s^{-1} \left \{ \frac{F(\Omega_0^{\phi}, \bar{w}_0)}{\sqrt{1 - \bar{\Omega}_{ls}^{\phi}}} \left [ 
\sqrt{a_{ls} + \frac{\Omega_0^r}{1 - \Omega_0^{\phi}}} - \sqrt{\frac{\Omega_0^r}{1 - \Omega_0^{\phi}}} \right ]^{-1} - 1 \right \}
\label{eq: la}
\end{equation}
with\,:

\begin{equation}
F(\Omega_0^{\phi}, \bar{w}_0) = \frac{1}{2} \int_{0}^{1}
{da \left [ a + \frac{\Omega_0^{\phi}}{1 - \Omega_0^{\phi}} a^{1 - 3\bar{w}_0} 
+ \frac{\Omega_0^r (1 - a)}{1 - \Omega_0^{\phi}} \right ]^{-1/2}} \ .
\label{eq: defeffe}
\end{equation}
In Eqs.(\ref{eq: la}) and (\ref{eq: defeffe}), $a_{ls} = (1 + z_{ls})^{-1}$ with $z_{ls}$ the redshift of last scattering that may be computed as \cite{HS96}\,:

\begin{equation}
z_{ls} = 1048 \left [ 1 + 0.00124 \left ( \Omega_{b,0} h^2 \right )^{-0.738} \right ] 
\left [ 1 + g_1 \left ( \Omega_{M,0} h^2 \right )^{g_2} \right ]
\label{eq: zls}
\end{equation}
with $(g_1, g_2)$ given in Ref.\,\cite{HS96}. The other quantities entering Eqs.(\ref{eq: la}) and (\ref{eq: defeffe}) are defined as follows \cite{DorLil01,DorLil02}\,:

\begin{equation}
\bar{c}_s = \frac{1}{\tau_{ls}} \int_{0}^{\tau_{ls}}{\left [ 3 + \frac{9}{4} \frac{\rho_b(\tau)}{\rho_r(\tau)} \right ]^{-2} d\tau} \ ,
\label{eq: defcs}
\end{equation}

\begin{equation}
\bar{w}_0 = \frac{\int_0^{\tau_0}{\Omega_{\phi}(\tau) w(\tau) d\tau}}{\int_0^{\tau_0}{\Omega_{\phi}(\tau) d\tau}} \ ,
\label{eq: wzmean}
\end{equation}

\begin{equation}
\bar{\Omega}_{ls}^{\phi} = \frac{1}{\tau_{ls}} \int_{0}^{\tau_{ls}}{\Omega_{\phi}(\tau) d\tau} \ ,
\label{eq: meanom}
\end{equation}
where $\tau = \int{a^{-1} dt}$ is the conformal time, $\rho_b$ and $\rho_r$ are the energy densities of the baryons and radiation respectively, $w(z)$ and $\Omega_{\phi} = \rho_{\phi}/\rho_{crit}(z)$ are the barotropic factor and the density parameter\footnote{Note that here we are using $\rho_{crit}(z) = 3 H^2(z)/8 \pi G$ rather than its present day value which has been adopted to get Fig.\,\ref{fig: ez}. As a consequence, in this section, we accordingly redefine $\Omega_{VdW}(z)$ as the ratio between $\rho_{VdW}(z)$ and $\rho_{crit}(z)$.} of the dark energy fluid. In order to use Eqs.(\ref{eq: la})\,-\,(\ref{eq: meanom}), we first remind that the only matter term in our model is the baryonic term so that we have to replace everywhere $\Omega_M$ with $\Omega_b$. On the other hand, the role of the scalar field fluid is played by the Van der Waals fluid so that all the quantities with the subscript $\phi$ have now to be evaluated using the Van der Waals energy density. Finally, the Hubble parameter is given by Eq.(\ref{eq: fried2}) with $\rho_{VdW}(z)$ obtained by solving Eq.(\ref{eq: etavsz}), while the present day value of the radiation density parameter is set as $\Omega_0^r = 9.89 \times 10^{-5}$. Note also that one has to choose a value for the index $n$ of the spectrum of primordial fluctuations which we set as $n = 1$.  

\begin{figure}[!t]
\begin{center}
\includegraphics{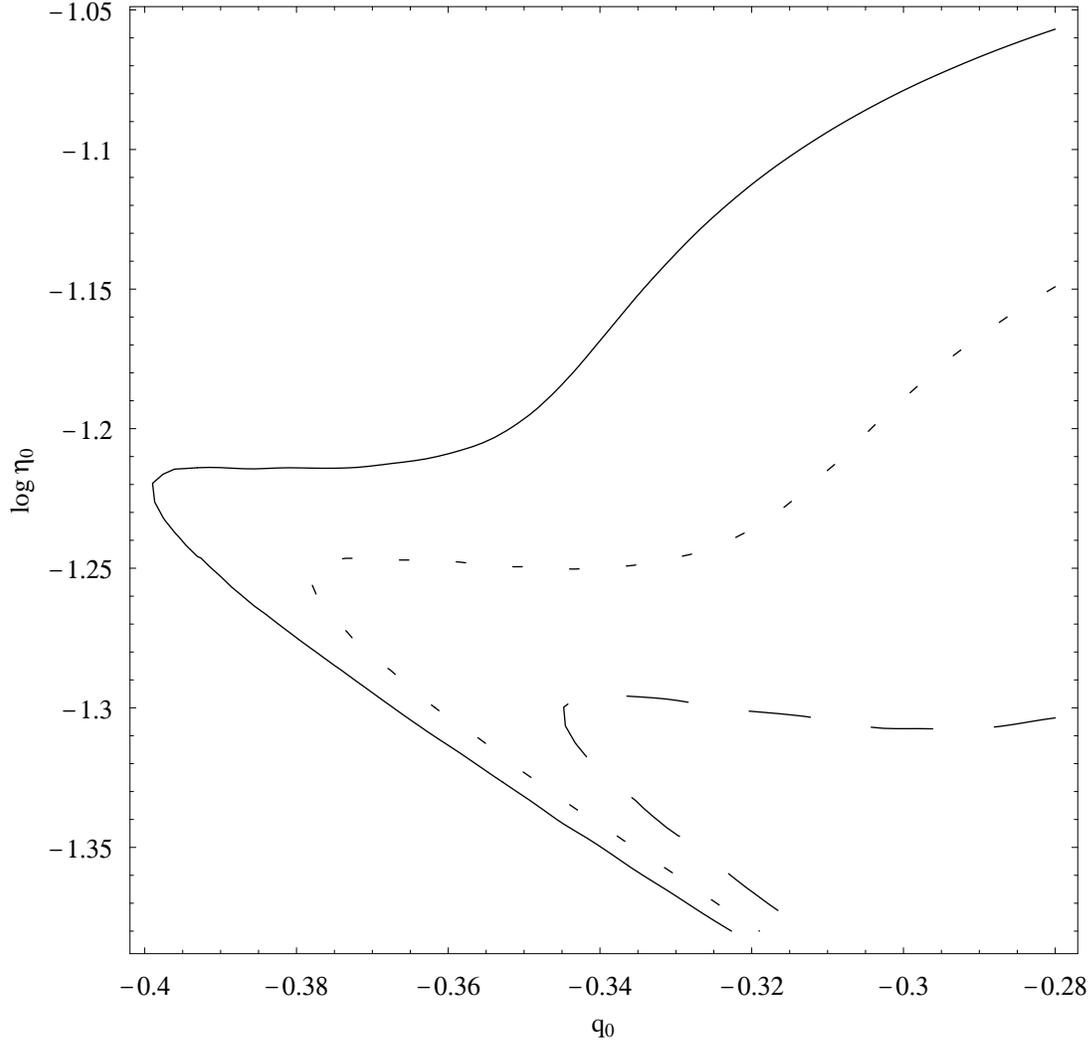}
\end{center}
\caption{Constraints in the $(q_0, \log{\eta_0})$ plane for models with $(h, \Omega_{b,0} h^2) = (0.664, 0.0214)$. The region to the right of the solid line individuates models that are consistent with $l_3^{Boom}$ within $3 \sigma$. Models with $(q_0, \log{\eta_0})$  belonging to the region of the parameter space to the right of the short and long dashed lines agree within $5 \sigma$ with $l_2^{WMAP}$ and $l_1^{WMAP}$ respectively. We only report the region of the parameter space delimited by the  $95\%$ CL constraints in Eq.(20).}
\label{fig: peaks}
\end{figure}

The position of the first two peaks in the CMBR anisotropy spectrum ha been determined with great accuracy by the WMAP measurements giving \cite{WMAP}\,:

\begin{equation}
l_1^{WMAP} = 220.1 \pm 0.08 \ \ , \ \ l_2^{WMAP} = 546 \pm 10 \ \ ,
\label{eq: l12obs}
\end{equation}
while the position of the third peak is more uncertain and may be estimated as \cite{Boom}\,:

\begin{equation}
l_3^{Boom} = 851 \pm 31 \ .
\label{eq: l3obs}
\end{equation}
Fig.\,\ref{fig: peaks} shows the region of the parameter space $(q_0, \log{\eta_0})$ that are consistent with the bounds from the position of the peaks. Note that the model predicted $l_m$ are affected by unknown errors due to both the approximated nature of the procedure adopted\footnote{To test the validity of the method adopted to compute $l_m$, one should compare the predicted values with what is obtained running a code that evaluates the full spectrum of perturbations. This test has been indeed performed with success only for scalar field dark energy with different choices of the self interaction potential. Although it is likely that the method works with the same goodness also in our case given the similarity of our model with the cosmological constant in the prerecombination epoch, a detailed comparison is still needed.} and to some subtleties we discuss later. To qualitatively take into account these systematics, we generously enlarge the error bars on the measured $l_m$ considering the $5 \sigma$ ($3 \sigma$) confidence ranges for WMAP (BOOMERanG) observed quantities. Considering the most stringent cut (that on $l_1$), Fig.\,\ref{fig: peaks} shows that it is indeed possible to find out models that are in agreement with both the fit to the dimensionless coordinate distances and the position of the first three peaks. However, some caution is needed. 

The method adopted to compute $l_m$ implicitly assume that the standard theory of perturbation may be applied to the Van der Waals quintessence scenario we are discussing. In this case, the sound speed $c_s$ entering the growth of perturbations is usually defined as  $c_s^2/c^2 = \dot{p}/\dot{\rho}$ with $c$ the speed of light. Using Eqs.(\ref{eq: vdweq}), we easily get\,:

\begin{equation}
\left ( \frac{c_s}{c} \right )^2 = \frac{9 \gamma (4 - \eta) (1 - \eta)^2}{4 (3 - \eta)^2}
\label{eq: cs}
\end{equation}
that at $z = 0$ reduces to\,:

\begin{equation}
\left ( \frac{c_s}{c} \right )_0^2 = \frac{6 (\eta_0 - 4) (\eta_0 - 1)^2 w_0}{(\eta_0 - 3) (3 \eta_0^2 - 9 \eta_0 + 8)}
\label{eq: cszero}
\end{equation} 
having used Eq.(\ref{eq: gammavsqz}). This quantity is not well behaved since, for the model parameters in the range determined above, may become negative giving rise to a formally imaginary sound speed. Moreover, during the evolution of the universe, $c_s^2$ can take negative values over a large redshift range. From this point of view, Van der Waals quintessence is similar to other unified dark energy models. In particular, Sandvik et al. \cite{Sandvik} have considered the particular case of the generalized Chaplygin gas and shown that it gives rise to oscillations or exponential blowup in the dark matter power spectrum inconsistent with observations. Although only the generalized Chaplygin gas has been investigated, they argue that similar problems also take place for every unified dark energy model because of these models having a negative $c_s^2$. This could be an evidence against the Van der Waals scenario we are proposing. Actually, things get different. First, one has to remind that $\dot{p}/\dot{\rho}$ is the {\it adiabatic sound speed}. If the fluid fluctuations were adiabatic, then the pressure perturbation $\delta p$ corresponding to a perturbation $\delta \rho$ in the energy density should be \cite{Hu}\,:

\begin{equation}
\delta p = (\dot{p}/\dot{\rho}) \delta \rho
\label{eq: adper}
\end{equation}
so that if $\dot{p}/\dot{\rho}$ becomes negative or singular the perturbations go unstable. This is the case for our model, but this problem may be avoided by resorting to non adiabatic pressure perturbation. In this case, the relation between $\delta p$  and $\delta \rho$ is generalized as\,:

\begin{equation}
\delta p = (\dot{p}/\dot{\rho}) \delta \rho + p \Gamma
\label{eq: dpdrho}
\end{equation}
with $\Gamma$ the entropy contribution related to the anisotropic scalar stress $\pi_L$. Using the freedom in the choice of $\pi_L$, it is then possible to define an effective sound speed that determines the scale under which the fluid energy is effectively smooth through the sound horizon. Well above this scale, stress gradients are negligible and the standard theory of perturbations applies. It is thus interesting to investigate how it is possible to implement such a mechanism for the Van der Waals quintessence and then study the problem of structure formation in this generalized scenario. We will address this topic in a forthcoming paper \cite{Moki}. 

\section{Conclusions}

The increasing bulk of astrophysical data accumulated in recent years has delineated a new standard cosmological paradigm. According to this picture, the universe is spatially flat and driven by an unknown form of dark energy leading to an accelerated expansion. Soon after the establishment of such a scenario, the hunt for candidates to the dark energy throne has started leading to the proposal of a plethora of mechanisms ranging from the old cosmological constant to various scalar field quintessence, to modification of Friedmann equations (motivated by extradimensions and braneworld theories) and higher order geometrical terms in the gravity Lagrangian \cite{fR1,fR2,fR3}. Although being completely different in their dynamical properties and underlying physics, they all share the ability of well fitting the same set of data. 

Furthermore, it is worth noting that the dark side of the universe is not only populated by dark energy, but also by dark matter whose nature is still far to be understood. It is thus tempting to ask whether these two ingredients are indeed different substances or two aspects of a single fluid whose properties are different from what we usually expect. To this regard, it makes sense wondering if the dust approximation used for matter is the right one to describe the equation of state of the dark matter. In a tentative to explain the observable quantities of our universe with the minimal number of ingredients, here we have investigated the possibility that dark matter and dark energy are actually a single fluid whose equation of state is that of a Van der Waals gas. In Van der Waals quintessence scenario, there is a single fluid whose equation of state comes directly from classical thermodynamics since the perfect gas approximation cannot be used to describe phase transitions which occur during the evolution of the universe. Although their contribution to the energy budget is nowaday subdominant, we also include baryons in the model since, because of the nonlinear character of the dynamical equations, they play an important role to determine how the main quantities evolve with the redshift.

Any theory, as elegant and motivated it can be, is meaningless if it is unable to give a coherent description of the universe as it is observed. That is why we have tested Van der Waals quintessence against the dimensionless coordinate distance to SNeIa and radio galaxies. This allows to narrow the parameter space of the model considering the present day values of the deceleration parameter and of the ratio between the critical density of the universe and the Van der Waals critical density. We find\,:

\begin{eqnarray}
-0.52 \le q_0 \le -0.33 \ \ , \ \ -0.42 \le \log{\eta_0} \le -0.11 \ \ & \ \ {\rm at \ 68\% \ CL} \ \ , \nonumber \\ 
~ \nonumber \\
-0.58 \le q_0 \le -0.28 \ \ , \ \ -1.38 \le \log{\eta_0} \le -0.05 \ \ & \ \ {\rm at \ 95\% \ CL} \ \ , \nonumber
\end{eqnarray}
with $(q_0, \log{\eta_0}) = (-0.38, -0.21)$ as best fit parameters. The constraints on $\log{\eta_0}$ are weak because of a serious degeneracy among the model parameters. Indeed, different models of this class predicts values of $y(z)$ which are in agreement with each other within the observational errors. However, lowering the uncertainties on $y(z)$ in the redshift range $(1.0, 2.0)$ or extending observations to higher redshifts (up to $z \sim 3 - 4$) will allow to break this degeneracy. Both these possibilities are likely to become realistic in a near future thanks to the next\,-\,to\,-\,come satellite and ground based experiments. 

Motivated by this encouraging result, we have compared the predicted age of the universe with that estimated from globular clusters in a model independent way. Unfortunately, it turns out that only a quite small region of the parameter space delimited by the above constraints is allowed by this test. Moreover, $t_0$ turns out to be systematically high so that one could argue against the proposed model. Actually, this conclusion is premature since our ignorance of the thermodynamical properties of the dark fluid makes it impossible to determine the redshift range over which Eq.(\ref{eq: vdwour}) holds. Indeed, during the evolution of the universe, the dark fluid may undergo phase transitions thus changing its equation of state in a unpredictable way. Since $t_0$ depends on the full evolutionary history of the universe, using Eq.(\ref{eq: vdwour}) at all $z$ may induce a systematic error if the relation $p = p(\rho)$ is altered. Given these theoretical uncertainties, overemphasizing the result of the age test should be avoided. 

As a further test, we have also evaluated the position of the first three peaks of the CMBR anisotropy spectrum using the analytical approximation developed in Refs.\,\cite{DorLil01,DorLil02}. Comparing with the measured $l_m$, we have individuated a region of the parameter space $(q_0, \log{\eta_0})$ which is consistent with both the distance fit and the peaks position. Although very encouraging, this result should be confirmed by evaluating the full spectrum and not the peaks position only. To this aim, one has to first investigate in more detail how perturbations grow in the Van der Waals quintessence scenario. Because of the negative adiabatic sound speed, perturbations go unstable and one has to take into account also the role of anisotropic stresses and entropy production. 

We would like to conclude with a general comment. Van der Waals quintessence has turned out to be an interesting scenario for describing the late universe and seems able to solve the puzzle of dark energy without adding exotic fluids or arbitrary modifications of Friedmann equations. On the other hand, classical thermodynamics tells us that the Van der Waals equation of state is only an approximated description of a realistic fluid. In our opinion, before invoking the help of {\it new physics}, it is worth wondering whether {\it classical physics} could still suggest the way to shed light on the dark side of the universe.

\ack

It is a pleasure to thank our friend Monica Capone for the fruitful discussions on this topic and Ester Piedipalumbo and Crescenzo Tortora for help with the position of the peaks. We are also grateful to an unknown referee for his stimulating report.

\end{document}